\documentstyle[12pt,titlepage,epsf]{article} 
\newcommand{\ba}{\begin{array}}
\newcommand{\ea}{\end{array}}
\newcommand{\bd}{\begin{displaymath}}
\newcommand{\ed}{\end{displaymath}}
\newcommand{\be}{\begin{equation}}
\newcommand{\ee}{\end{equation}}
\newcommand{\bea}{\begin{eqnarray}}
\newcommand{\eea}{\end{eqnarray}}

\begin{document}
\setcounter{page}{0}
\thispagestyle{empty}

\begin{flushright}
BU-TH/96-1 \\ IISc-CTS-8/96\\ MRI-PHY/6/96\\ {\bf hep-ph/9605407} \\
\end{flushright}

\begin{center}
{\large\bf UNUSUAL CHARGED HIGGS SIGNALS AT LEP-2 \\[0.5truein]}
{Dilip Kumar Ghosh$^1$, Rohini M. Godbole$^2$ and 
Biswarup Mukhopadhyaya$^3$\\}
\end{center}

\vskip 30pt

\begin{center}
{\large\bf ABSTRACT}\\
\end{center}

\noindent
We have made a detailed study of the signals produced at LEP-2
from charged scalar bosons whose dominant decay channels are
into four fermions.  The event rates as well as kinematics of
the final states are discussed when such scalars are either
pair-produced or are generated through a tree-level interaction
involving a charged scalar, the $W$ and the $Z$. The backgrounds
in both cases are discussed. We also  suggest the possibility of
reconstructing the mass of such a scalar at LEP-2.

 
\vskip 40pt

\footnotesize
\noindent
$^{1}$ Department of Physics, University of Bombay,
Vidyanagari,Santa Cruz (East), Mumbai 400 098,  India.  E-mail:
dghosh@theory.tifr.res.in  \\ $^{2}$ Centre for Theoretical
Studies, Indian Institute of Science, Bangalore - 560 012,
India. \\ (On leave of absence from  Department of Physics,
University of Bombay,India).\\ E-mail: rohini@cts.iisc.ernet.in
\\ $^{3}$ Mehta Research Institute, 10 Kasturba Gandhi Marg,
Allahabad - 211 002, India. \\ E-mail: biswarup@mri.ernet.in
\normalsize

\newpage
\textheight=8.9in

\section{INTRODUCTION}

A major part of the activities in high energy physics today
concerns the signatures of the Higgs boson which is responsible
for electroweak symmetry breaking in the Glashow-Salam-Weinberg
model, popularly known as the standard model (SM). The SM
symmetry breaking sector contains a complex scalar doublet, from
which one neutral physical state ultimately remains while the
others are gauged away to give masses to the weak vector bosons.
This scalar decays primarily into either a pair of vector bosons
or into a fermion-antifermion pair, depending on its mass.  A
number of search strategies  have accordingly been formulated at
both electron-positron and hadronic colliders \cite{hhg} for
this scalar.

There is, however, a parallel stream of activities, going on for
quite some time now, to find out whether there is some new
physics beyond the standard model. One aspect of this type of
effort is to see the phenomenological implications of a larger
scalar sector. This includes the possibility of models with two
or more Higgs doublets. Also, scenarios with higher scalar
representations of SU(2) cannot  {\it a priori} be ruled out. In
fact, some such representations could be theoretically
advantageous for example, triplet scalars can give us a natural
explanation of left-handed Majorana masses for neutrinos
\cite{mn}. The important thing to note is that in certain
situations, the dominant decay modes of some physical scalars
could be different from those predicted by the SM.  If these
happen to be the lightest scalar states, then our first
encounter with elementary spin-zero particles may very well be
through some of these new channels. For example, as we shall see
below, there is a distinct possibility of scalars with no
tree-level fermion-antifermion interaction \cite{phob}.  Such
scalars, in a certain mass range, may decay dominantly into four
fermions \cite{gmn}.  Also, some of them may be produced
\cite{l1} through a tree-level interaction involving a charged
scalar, the $W$ and the $Z$, something that is disallowed in the
SM and its extensions into two or more Higgs doublets. In this
paper, we discuss some of these unusual signals of a charged
scalar at LEP-2.

In section 2 we outline the theoretical scenarios in which these
non-standard signals of a scalar may be observed. Sections 3 and
4 are devoted respectively to the signals of pair-produced
charged scalars and those produced through the
$H^{\pm}W^{\mp}Z$-coupling. We conclude the discussion in
section 5.
  
\section{THEORETICAL MOTIVATIONS}

As we have already stated, there are two crucial components in
the unusual interactions of a scalar, namely $(i)$ the absence
of  tree-level fermion-antifermion interactions, and $(ii)$ the
presence of  a tree-level $H^{\pm}W^{\mp}Z$-coupling. We give
below some specific examples of models where one or both of the
above features are present. For further details of these models,
we refer the reader to the existing literature.

\noindent
$(a)$ A two Higgs doublet model can be constructed \cite{hks}
where only one doublet couples to fermions(f) with isospin both
$\frac{1}{2}$ and -$\frac{1}{2}$.  This is one of the options
available to ensure that there is no flavour-changing Yukawa
coupling at the tree-level \cite{nfc}. In this case, in general,
physical states are formed after the two doublets mix. It can be
shown that one of the two neutral physical scalars ceases to
have any $f\bar{f}$-coupling if the mixing angle is either 0 or
$\pi /2$ (caused by some discrete symmetry in the potential).
But the charged scalar will still have a fermionic coupling. On
the other hand, if there are more than two doublets, (for
example, those in the Weinberg model of CP-violation)
\cite{wei}, charged Higgs physical states totally decoupled from
fermions can be envisioned. However, none of these doublet
extensions allows a tree-level $H^{\pm}W^{\mp}Z$-interaction.

\noindent
$(b)$ A model with a Higgs doublet together with a complex
($Y=2$) triplet has been proposed to explain the origin of
left-handed Majorana masses for neutrinos.  Such a Higgs
structure can be built within a left-right symmetric scenario
\cite{mn}.  The introduction of a triplet makes it possible to
have a tree-level $H^{\pm}W^{\mp}Z$-interaction, although the
physical charged Higgs state need not in general be decoupled
from fermion-antifermion pairs \cite{comp}.  It should be noted
here that all triplet (and higher) representations imply a
departure from  unity in the tree-level value of the
$\rho$-parameter, defined as $\rho = {\frac{m_W^2}{m_Z^2 cos{^2}
\theta_W}}$. With just one complex triplet, such a contradiction
with experiments \cite{pdg} can be avoided by assuming that the
vacuum expectation value (VEV) of the neutral member of the
triplet is very small compared to the doublet VEV, as a result
of which the triplet contributes negligibly to the $W$-and
$Z$-masses. Such a solution, however requires fine tuning.

\noindent
$(c)$ A more aesthetic  way out of the problem with $\rho$ is to
postulate \cite{gg} that there is more than one triplet (or
higher representation), arranged in such a manner that their
{\it net contribution} to $\rho$ is 1. One way to do this is to
have one complex ($Y=2$) and one real ($Y=0$) triplet along with
the conventional doublet. Moreover, one has to assume that the
VEV's of the neutral components of the two triplets are equal,
as a consequence of a custodial symmetry. This allows the
triplet to contribute substantially to the gauge boson masses
without affecting the value of $\rho$. It has been shown
\cite{gold} that the scalar potential can be constructed to obey
such a custodial symmetry upto arbitrary orders, although it
requires fine-tuning (to the same extent as that in the SM
itself \cite{gun1}) to maintain the symmetry when gauge
couplings are switched on. This scenario also implies a
tree-level $H^{\pm}W^{\mp}Z$ vertex. In addition, it leads to a
5-plet of scalar physical states ($H^{\pm \pm}, H^{\pm}, H^{0}$)
under the $SU(2)$ custodial symmetry which have no overlap with
the doublet, and hence have no fermion-antifermion couplings at
all \cite{gun2}.
                                                  
Among the above examples, ($a$) has feature ($i$) only, ($b$)
has feature ($ii$) only, and ($c$) has both. In the remaining
discussion, we shall not make any specific model assumption.
However, we shall consider the signals of a charged Higgs which
has both of the above features.

Here we are concerned with signals at the LEP-2. Consequently
our attention will be focussed on a charged scalar of mass
between 45 and 80 GeV approximately. There are two ways in which
such an object can be formed in $e^{+} e^{-}$ annihilation,
namely, pair production,  via $s$ channel $\gamma$ and $Z$
exchange and associated production via the
$H^{\pm}W^{\mp}Z$-interaction. The second possibility has
already been studied \cite{gmn,pil}, although \cite{pil}
discusses it more in the context of the next linear collider
(NLC). In what follows, we will give details of the signals in
the first case, and will do some further analysis of the signals
vis-a-vis backgrounds in the second case also.  Being in the
aforesaid mass range, the charged scalar can decay either into
four fermions at the tree-level, mediated by a virtual $W$ and a
virtual $Z$, or into a fermion-antifermion pair at the one-loop
level.  The relative strengths of these two channels are
independent of the strength of the $H^{\pm}W^{\mp}Z$ coupling,
which occurs identically in both decay widths.  It has  already
been shown \cite{gmn}, contrary to earlier claims, that the
tree-level process is dominant once the Higgs is as massive as
about 50 GeV.  Therefore, for such masses as can be looked for
at LEP-2, we assume a 100\% branching ratio for the four-fermion
decay of a singly charged scalar in our analysis.\footnote {In
principle, if there are triplets, a singly charged scalar can
decay into two leptons if lepton number violation is allowed.
We disregard that possibility here.}

\section{PAIR-PRODUCTION OF CHARGED SCALARS}

A charged scalar can be pair-produced in $e^{+}e^{-}$
annihilation through $s$-channel diagrams mediated by the $Z$
and the $\gamma$. The production cross-section is
model-independent \cite{kom}. In Figure 1 we show this
cross-section as a function of the charged Higgs mass for
centre-of-mass energies 176 GeV and 192 GeV respectively. The
corresponding cross-sections for associated charged Higgs
production via the $H^{\pm}W^{\mp}Z$ vertex are also shown.  We
can see from the Figure that the latter can be quite sizable.
However, since these numbers are for maximal mixing between the
doublet and exotic Higgs representation(s), this mechanism
dominates over pair-production only if the mixing angle is
large. This is allowed only  in option ($c$) discussed in
section 2.

For the charged scalars of our interest, each can decay (via a
virtual $W$ and a $Z$) into four fermions, which can be ($a$)
two quark-antiquark pairs($q \bar qq \bar q$), ($b$) a
quark-antiquark and a lepton-antilepton pair($q \bar q~
\ell^{+}\ell^{-}$), and ($c$) two lepton-antilepton
pairs($\ell^{+}\ell^{-}\ell^{+}\ell^{-}$). Though the final
states with increasingly higher number of leptons have  cleaner
signatures, one pays a heavy price in terms of branching ratios.
When two charged scalars are involved, one can hope to get a
sufficient number of events only by studying the final states
with the largest branching ratios. For this, at least one of
them has to decay purely hadronically. The other will either go
to four jets, or to two jets (via the virtual Z) and a lepton
and a neutrino (via the virtual $W$).  While the former of these
has a total branching ratio of about 25\%, that for the latter
case is approximately 14\% (including leptons of both positive
and negative charges). It is straightforward to see that all the
other final states will give experimentally insignificant rates.
(By `leptons' here we mean electrons and muons).

In Figures 2($a,b$) we present the multijet cross-sections,
calculated in a parton level Monte Carlo simulation, as
functions of the charged Higgs mass for a centre-of-mass energy
of 176 GeV. In Figure2($c$) we present the same thing but at  a
higher centre-of-mass energy of 192 GeV. We have used a
jet-merger criterion that two jets are merged into one if
$\Delta R \leq 0.7$, where $\Delta R^2 = \Delta \phi^2 + \Delta
\eta^2$, $\Delta \phi$ and $\Delta \eta$ being the separations
in azimuthal angle and pseudorapidity of the initial partons.
We find that even though the final state nominally has 8
partons, it has jet multiplicities 6-8 as a result of
jet-merging. As we can see from the Figure, the largest
cross-section is in the 6-jet final state irrespective of the
Higgs masses and centre-of-mass energies considered here.  One
should note that in the Figure 2($a$) cross-sections have been
obtained  without using any invariant mass cut, whereas in the
Figure 2($b$) and Figure 2($c$) we have demanded that the
invariant mass of no combination of jets lies within $m_W \pm 5$
GeV. We will explain the need for such a cut later.

In Figure 3 we show the distribution in $\Delta R$ between
different jet pairs before merging.  This shows that the jets
coming out of the two oppositely charged scalars are clearly
separated into two hemispheres; if one labels the jets from one
scalar by the numbers 1 to 4 and those from the other by 5 to 8,
then $\Delta R$ within the same set tends to peak between 0.5
and 1, while the peak between opposite sets is observed around
3.  As we can see from this Figure, if one puts a judicious cut
on the maximum value of $\Delta R$, one can reduce contamination
of the decay products  of one of the scalars from those of the
other, making it possible to think about the measurement of the
scalar mass by reconstructing the invariant mass in a given
hemisphere.

Figure 2($a$) shows an initial rise of the cross-section with
$m_H$, particularly with high jet multiplicities, upto a certain
value. The 6-jet events dominate principally in the low Higgs
mass region. This is because a higher mass Higgs boson has lower
3-momentum, which  causes the four daughter fermions to be more
spherically spread out, thereby reducing the probability of
merging. This gives rise to more 7- and 8-jet events for a
heavier Higgs until phase space suppression of the cross-section
takes over. In contrast to Figure 2($a$), in Figures 2($b,c$)
where the invariant mass cut has been used, the 6-jet events
dominate over 7 and 8 jet events for the complete range of Higgs
masses we are interested in. This is because of the fact that in
the 6-jet final state, merging of jets increases the total 6-jet
invariant mass, so the peak around ${m_W}\pm 5$ GeV is smeared
out. The energy distributions for the four most energetic jets
(after jet merging) have also been shown in Figure 4.
  
Backgrounds to the above events are most serious for 6 jets. As
the number of jets increases, the direct QCD final states become
suppressed. However, the strongest SM backgrounds in this case
are 6-jets coming from $W^{\pm}$-pairs produced at LEP-2. These
can be as much as 2 to 5 times the signals for, say, $m_H = 55$
GeV, depending on the jet-merging criterion \cite{j6}.  To
remove this background, we have demanded in our parton level
Monte Carlo that no combination of jets may have invariant mass
in the range $m_W \pm 5$ GeV on both sides.  The 7- and 8-jet
events are relatively immune to such backgrounds. On the whole,
for $\sqrt{s} = 192$ GeV and $m_H \leq 60$ GeV there can be
still some 7-odd events even with 7-jets, and upto 15 events
with 6-jets (assuming an integrated luminosity of $500$pb$^{-1}$
per year),  provided that 7 jets can be identified, this should
be a reasonable rate. For higher Higgs masses the rate starts
falling even further.

When one charged Higgs decays into a pair of jets together with
a lepton and $p_T\!\!\!\!\!\!/$~~, the final states become
relatively cleaner  from the viewpoint of backgrounds. However,
because of branching fractions one has a further suppression of
about ($\frac{14}{25}$), as we have mentioned earlier.  The
event topologies are similar to the case of all jets; a
separation into opposite hemispheres can still be seen for the
particles coming out of the two scalars.  We also impose a lower
cut of 2 GeV on the lepton energy throughout, which does not
really affect the signal as the lepton energy peaks beyond 10
GeV. The isolations between the charged lepton and the jets have
been shown in Figure 5; for a typical jet in the same hemisphere
the distribution peaks at a healthy value of $\Delta R \approx
1$, whereas for an oppositely directed jet the peak is around 3.
It is also to be noted that the lepton generally emerges at high
angles with the beam pipe, as its rapidity distribution shows in
Figure 6.

\section{SINGLY PRODUCED CHARGED SCALARS}

As has been mentioned already, an exotic charged Higgs may also
be produced at an $e^{+}e^{-}$ collider through a tree-level
$H^{\pm}W^{\pm}Z$ vertex.  This process has been investigated in
some earlier works. Of these, reference \cite{pil} gives details
of the expected event shapes and the SM backgrounds for an
$e^{+}e^{-}$ collider with centre-of-mass energy in the range
500 GeV-- 2 TeV.  Here we focus on them in the special context
of LEP-2, in supplement to our earlier study.

There are two channels of production a $Z$-mediated $s$-channel
graph producing a charged scalar together with a real or a
virtual $W$, or a $WZ$-fusion process producing a charged scalar
together with  ${\bar e} {\bar{\nu_e}}(e^{+}\nu_e)$.  The latter
diagram becomes important for $\sqrt{s} \sim 500$ GeV or above
because of s-channel suppression. At LEP-2, the former is found
to be play the dominant role. The production cross-section is
proportional to $\sin^2 \theta_H$, where $\theta_H$ is the
mixing angle of the doublet Higgs with that belonging to a
higher representation. In other words, it gives the fractional
contribution of the exotic scalar representation to the
$W$-mass. In models with a custodial symmetry of the type
described in section 2, $\sin\theta_H$ can be as high as of the
order of unity without any phenomenological contradiction.  The
results shown in the Figures corresponds to $\sin\theta_H = 1$;
for other values the rates can be obtained by suitable scaling.
Let us add here for the sake of completeness that in case (b)
described in section 2, LEP data impose the restriction $\sin^2
\theta_H < 0.009$.

As Figure 1 shows, this kind of production has higher
cross-sections compared to pair-production for larger values of
the scalar mass, provided that $\sin\theta_H$ is large. Again,
the event rates are maximum when out of the 6 fermions in the
final state, at least 4 are hadrons. In Figure 7 the
cross-sections are shown for the multijet signals, where, again,
the same jet-merging criterion has been employed. Although the
4-and 5-jet final states do not stand much of a chance of
detection against QCD backgrounds, it should be noted that the
6-jet events have higher rates than those in the previous
section so long as $\sin\theta_H$ is large. Remembering that the
invariant mass cuts are applied here, this seems to be a rather
optimistic situation for the exotic Higgs signals.  When either
the associated  $W$ or the virtual $W$ in Higgs-decay gives rise
to a pair of leptons, the events become even cleaner, although a
corresponding suppression of approximately ($\frac{2}{7}$) with
respect to the hadronic channel in the overall rate is
inevitable.

Reference \cite{pil} identifies the following SM processes as
major backgrounds to the $H^{\pm}W^{\mp}Z$-induced production:
($a$) $e^{+}e^{-} \longrightarrow W^{+}W^{-}Z$, ($b$)
$e^{+}e^{-} \longrightarrow e^{+}e^{-}W^{+}W^{-}$ and ($c$)
$e^{+}e^{-} \longrightarrow e^{+}W^{-}Z \nu$. However, these can
be serious only when the centre-of-mass energy is large enough
and the Higgs can actually decay into a real $W$ and a $Z$. For
the scalar mass range that we are concerned with at LEP-2,
background ($a$) is not kinematically allowed.  Also, so long as
$m_H \leq m_W$, ($b$) and ($c$) can be removed by demanding that
no fermion pair reconstructs to the $W$ or the $Z$-mass. This
includes both invariant mass reconstruction and, for leptons
with neutrinos, Jacobian peaks in the transverse plane. In a
similar way $e^{+}e^{-} \longrightarrow W^{+}W^{-} \gamma$ or
$e^{+}e^{-} \longrightarrow ZZ \gamma$ does not pose a serious
threat; moreover, a simple estimate shows their cross-sections
to be on the order of $10^{-3}$pb. Another potential source is
when a $Z$-pair is produced in $e^{+}e^{-}$ collision. One of
these $Z$'s may go to a pair of jets, while the other may go to
a $\tau^{\pm}$-pair. Further decays of the tau's may mimic the
$jets + lepton + p_T\!\!\!\!\!\!/$~~or the multijet signal.
However, without any cuts, the signal is higher than the
backgrounds by at least a factor of 2. In addition, the softness
of the products of tau-decay provides a way of suppressing them
by suitable kinematic cuts.

\section{CONCLUSIONS}

We have performed a detailed analysis of the types of events
expected at the LEP-2 if there happen to charged scalars in the
mass range of 50-80 GeV, which do not have tree-level
fermion-antifermion couplings.  On the whole, the rates of
single production from the $H^{\pm}W^{\mp}Z$-vertex are large
provided that the contribution of non-doublet VEV's to the gauge
boson masses is dominant. On the other hand, if that
contribution is small, then one has to depend primarily upon
pair-production which has smaller rates. But then the high
multiplicity of final state particles seem to help the signals
in rising above the backgrounds once appropriate cuts are
imposed.

\section{ACKNOWLEDGEMENTS}

The authors would like to thank S. Dutta, G. Majumdar, N. Parua,
S. Raychaudhuri and S. Sarkar for discussions.  BM wishes to
thank the Department of Physics, University of Bombay, for
hospitality while this work was in progress. DKG acknowledges
financial support from the University Grants Commission,
Government of India, and also the hospitality of the Mehta
Research Institute towards the concluding stage of the project.
The work of RMG is partially supported by a grant (no.
3(745)/94/EMR(II)) of the Council of Scientific and Industrial
Research, Government of India.

\newpage
\centerline {\large {\bf Figure Captions}}

\noindent{\sl Figure 1} : Total cross-section (1) for
$e^{+}e^{-} \rightarrow H^{\pm}W^{\mp}$ in pb ($a$) at
$\sqrt{s}=176$GeV and ($b$) at $\sqrt{s}=192$GeV; (2) for
$e^{+}e^{-}\rightarrow H^{+}H^{-}$ in pb ($c$) at
$\sqrt{s}=176$GeV and ($d$) at $\sqrt{s}=192$GeV as a function
of charged Higgs mass.

\bigskip

\noindent{\sl Figure 2($a$)} : Multijet cross-section (without
invariant mass cut as discussed in the text) for
$e^{+}e^{-}\rightarrow H^{+}H^{-}\rightarrow jets$ in pb at
$\sqrt{s}=176$GeV as a function of charged Higgs mass. ($a$)
6-jets, ($b$) 7-jets and ($c$) 8-jets in the final states
respectively.

\bigskip

\noindent{\sl Figure 2($b$)} : Multijet cross-section (with
invariant mass cut as discussed in the text) for
$e^{+}e^{-}\rightarrow H^{+}H^{-}\rightarrow jets$ in pb at
$\sqrt{s}=176$GeV as a function of charged Higgs mass. ($a$)
6-jets, ($b$) 7-jets and ($c$) 8-jets in the final states
respectively.

\bigskip

\noindent{\sl Figure 2($c$)} : Multijet cross-section (with
invariant mass cut as discussed in the text) for
$e^{+}e^{-}\rightarrow H^{+}H^{-}\rightarrow jets$ in pb at
$\sqrt{s}=192$GeV as a function of charged Higgs mass. ($a$)
6-jets, ($b$) 7-jets and ($c$) 8-jets in the final states
respectively.

\bigskip

\noindent{\sl Figure 3} : Isolation  between different jet pairs
for $e^{+}e^{-}\rightarrow H^{+}H^{-}\rightarrow jets$ at
$\sqrt{s}=192$GeV and $m_H=55$GeV. $\Delta R$ between two jets
from same hemisphere: ($a$) and ($b$); $\Delta R$ between two
jets from opposite hemispheres: ($c$) and ($d$).

\bigskip

\noindent{\sl Figure 4} : Jet energy distributions for the four
most energetic jets for $e^{+}e^{-}\rightarrow
H^{+}H^{-}\rightarrow jets$ at $\sqrt{s}=192$GeVand at
$m_H=55$GeV.

\bigskip

\noindent{\sl Figure 5} : Isolation  between different jet pairs
and between a charged lepton and jet for $e^{+}e^{-}\rightarrow
H^{+}H^{-}\rightarrow jets+\ell+\nu$ at $\sqrt{s}=192$GeV and
$m_H=55$GeV. $\Delta R$: between two jets($a$), jet and charged
lepton ($b$) from same hemisphere; and a jet and charged lepton
($c$) from opposite hemisphere.

\bigskip

\noindent{\sl Figure 6} :  Lepton rapidity distribution for
$e^{+}e^{-}\rightarrow H^{+}H^{-}\rightarrow jets+\ell+\nu$ at
$\sqrt{s}=192$GeV and $m_H=55$GeV.

\bigskip

\noindent{\sl Figure 7} : Multijet cross-section (with invariant
mass cut) for $e^{+}e^{-}\rightarrow H^{\pm}W^{\mp}\rightarrow
jets$ in pb at $\sqrt{s}=192$GeV as a function of charged Higgs
mass. ($a$) 5-jets, ($b$) 6-jets and ($c$) 4-jets in the final
states respectively.

\bigskip
\newpage

\begin{figure}[htb]
\epsffile[100 390 120 700]{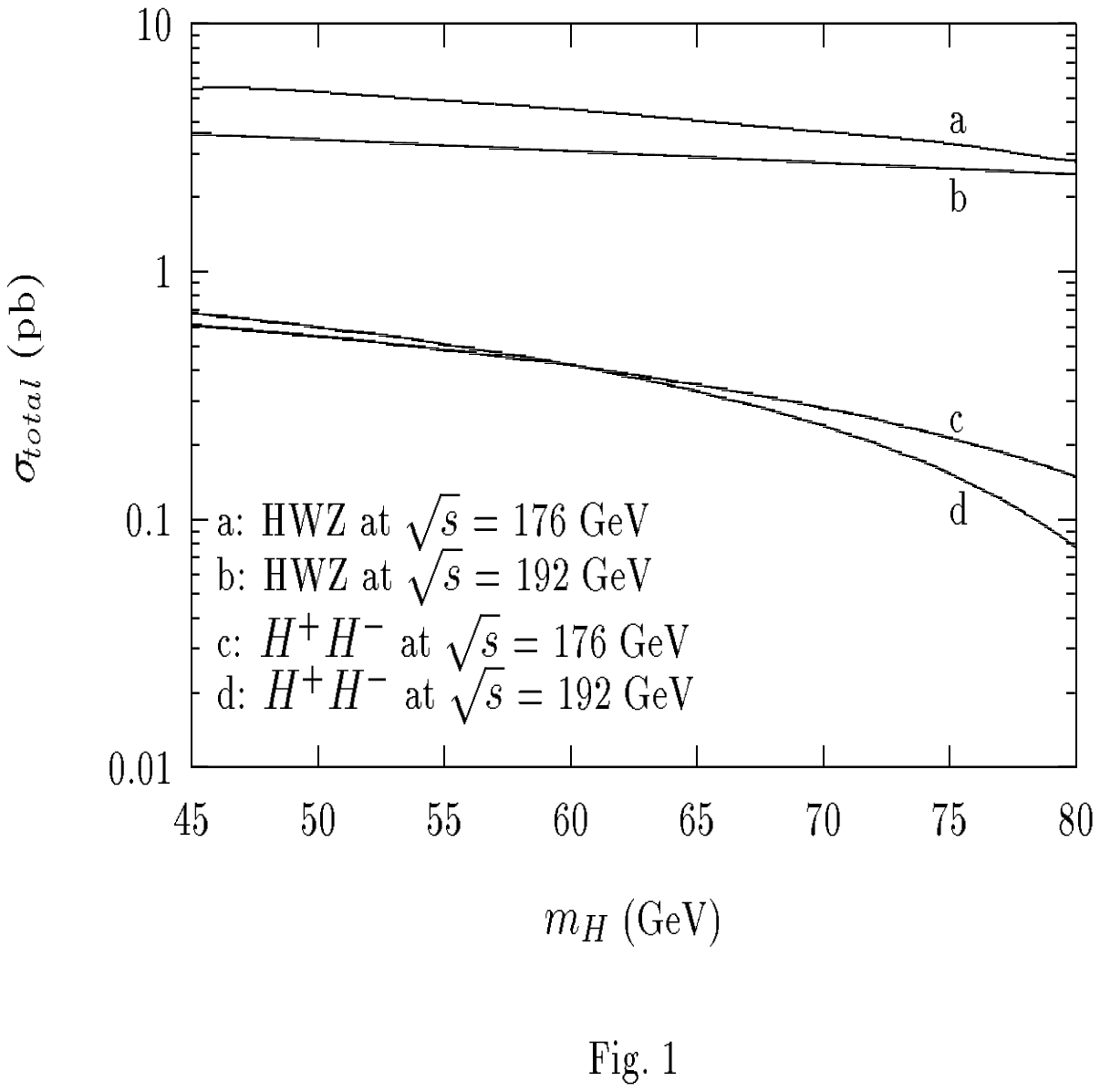}
\end{figure}

\newpage

\begin{figure}[htb]
\epsffile[100 390 120 700]{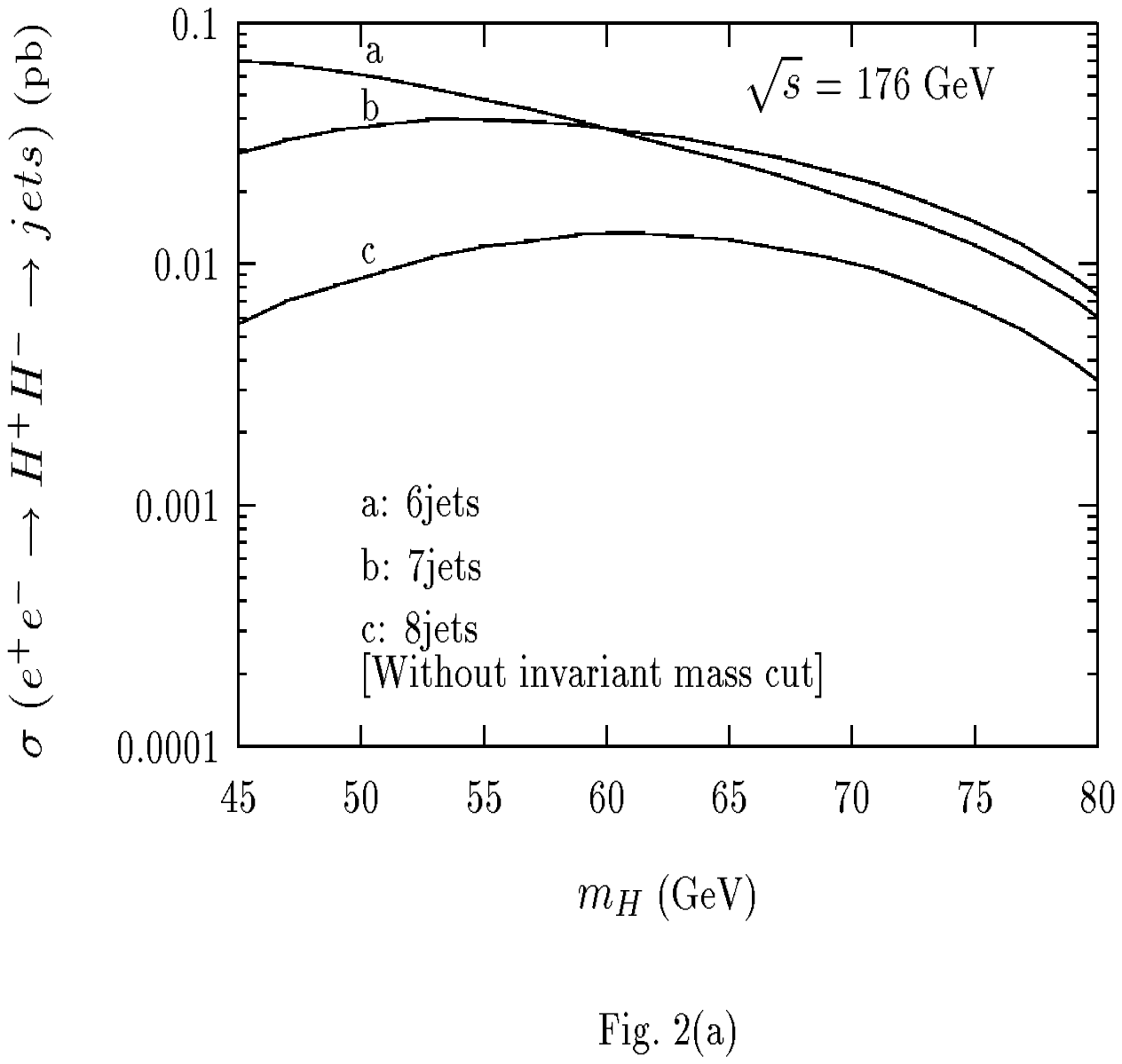}
\end{figure}

\newpage

\begin{figure}[htb]
\epsffile[100 390 120 700]{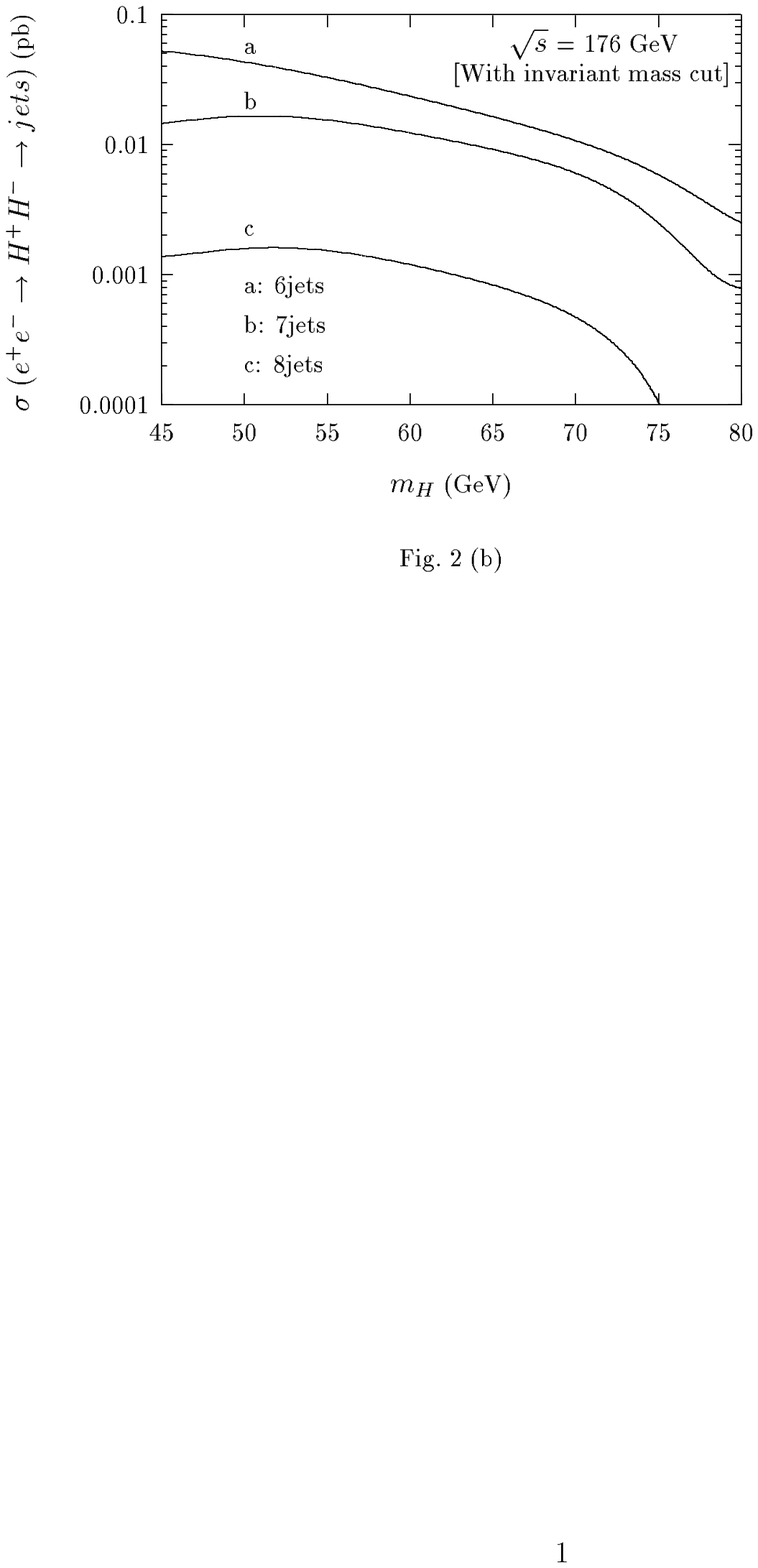}
\end{figure}

\newpage

\begin{figure}[htb]
\epsffile[100 390 120 700]{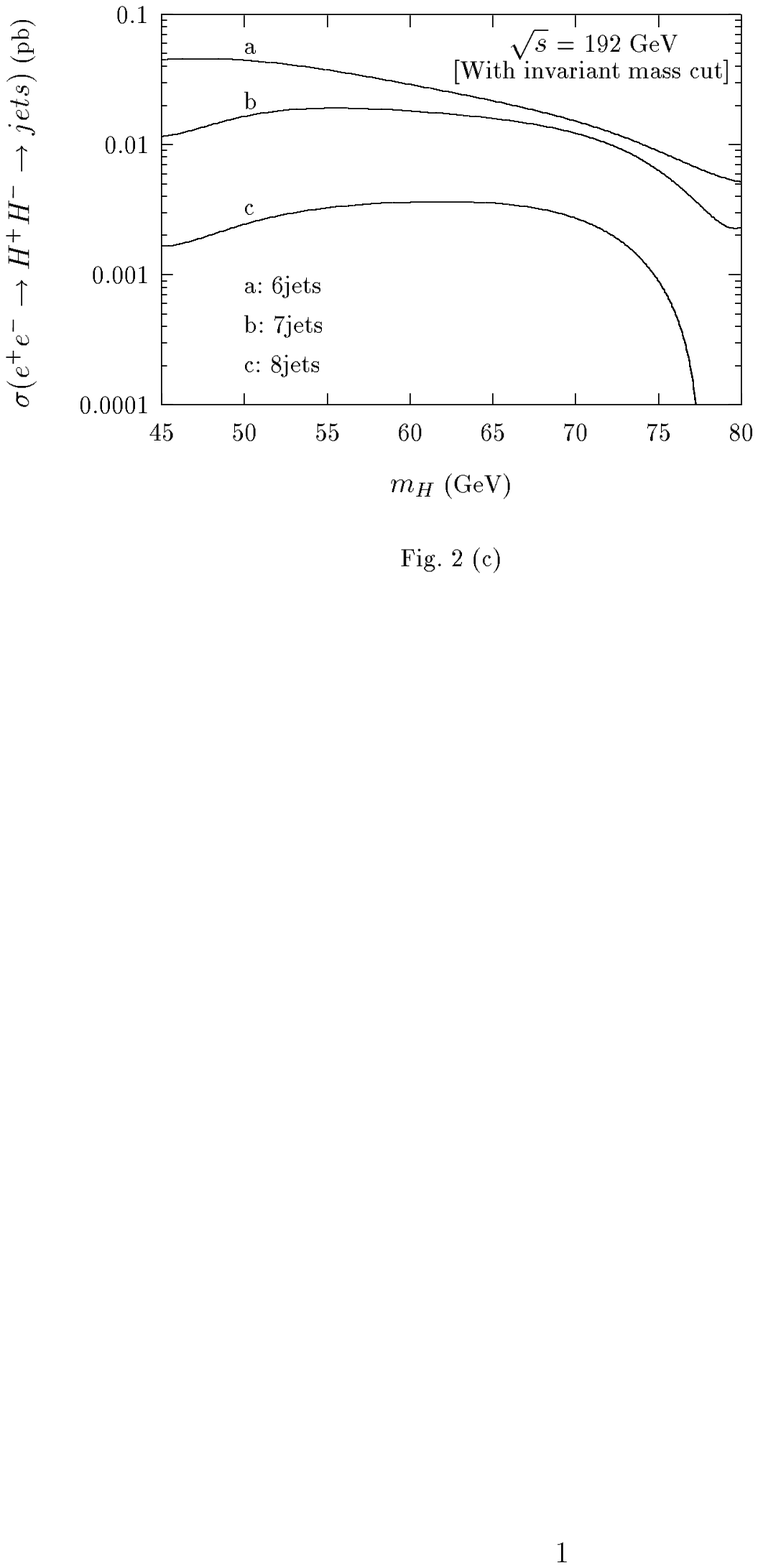}
\end{figure}

\newpage

\begin{figure}[htb]
\epsffile[100 390 120 700]{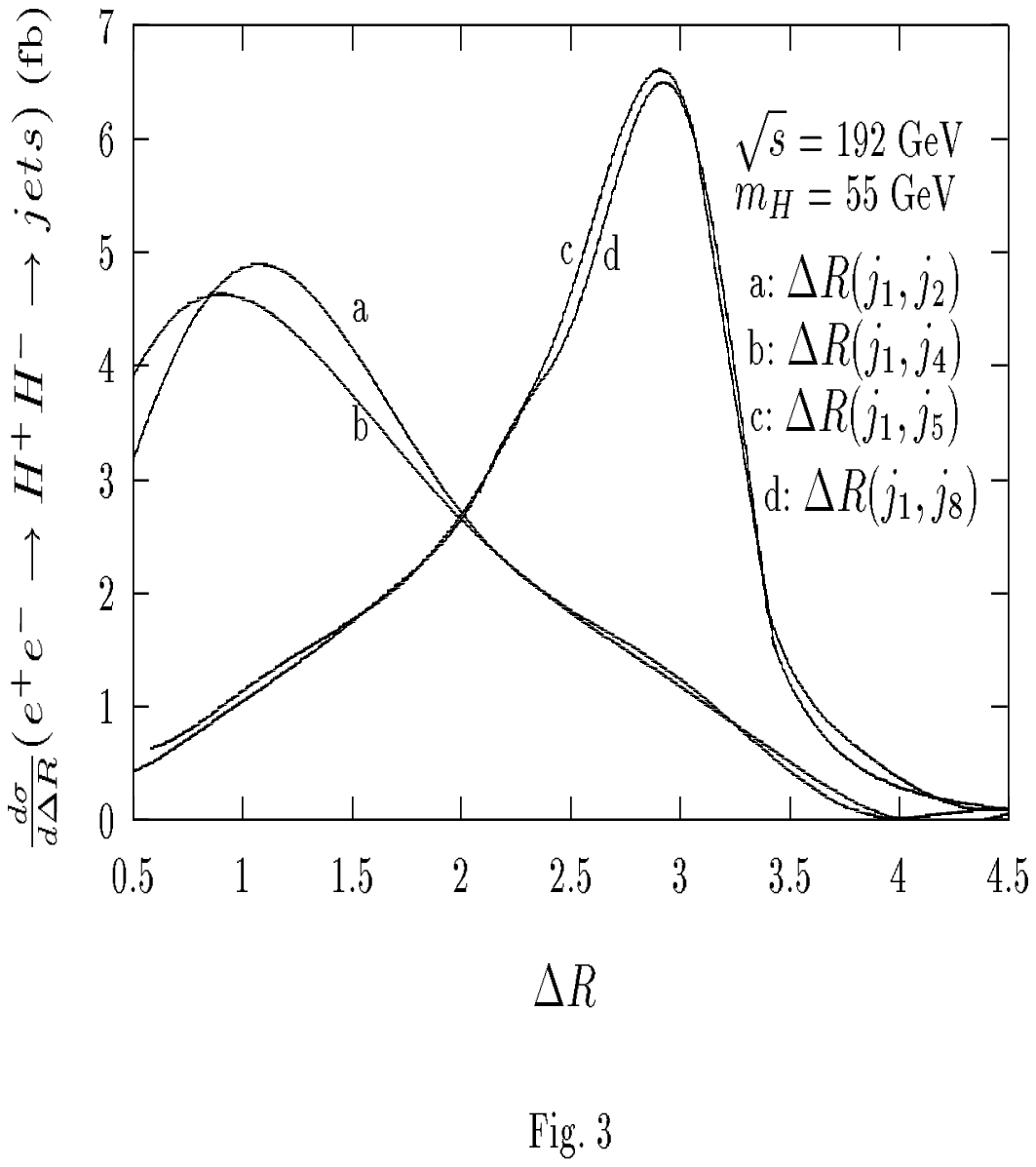}
\end{figure}

\newpage

\begin{figure}[htb]
\epsffile[100 390 120 700]{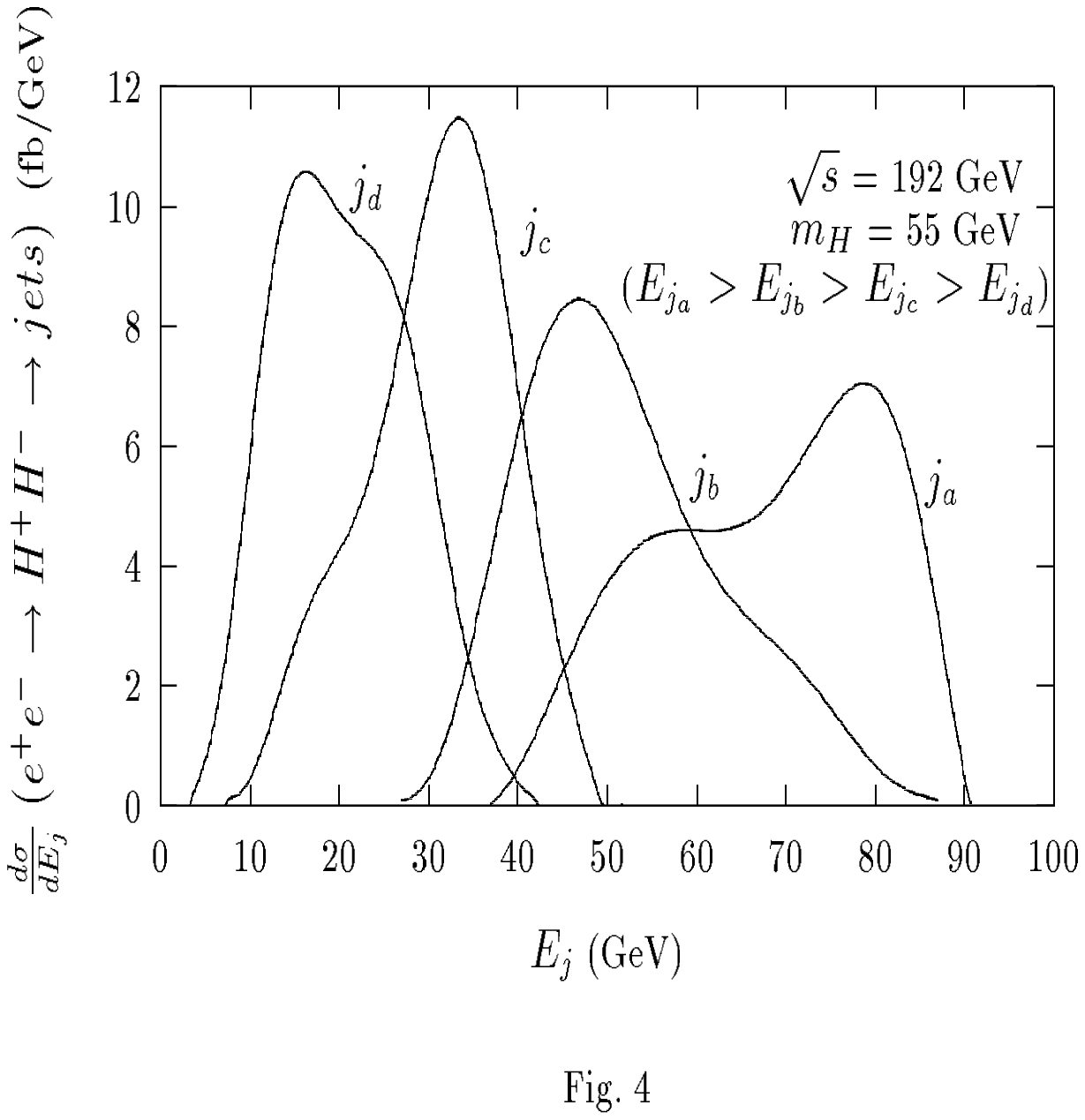}
\end{figure}

\newpage

\begin{figure}[htb]
\epsffile[100 390 120 700]{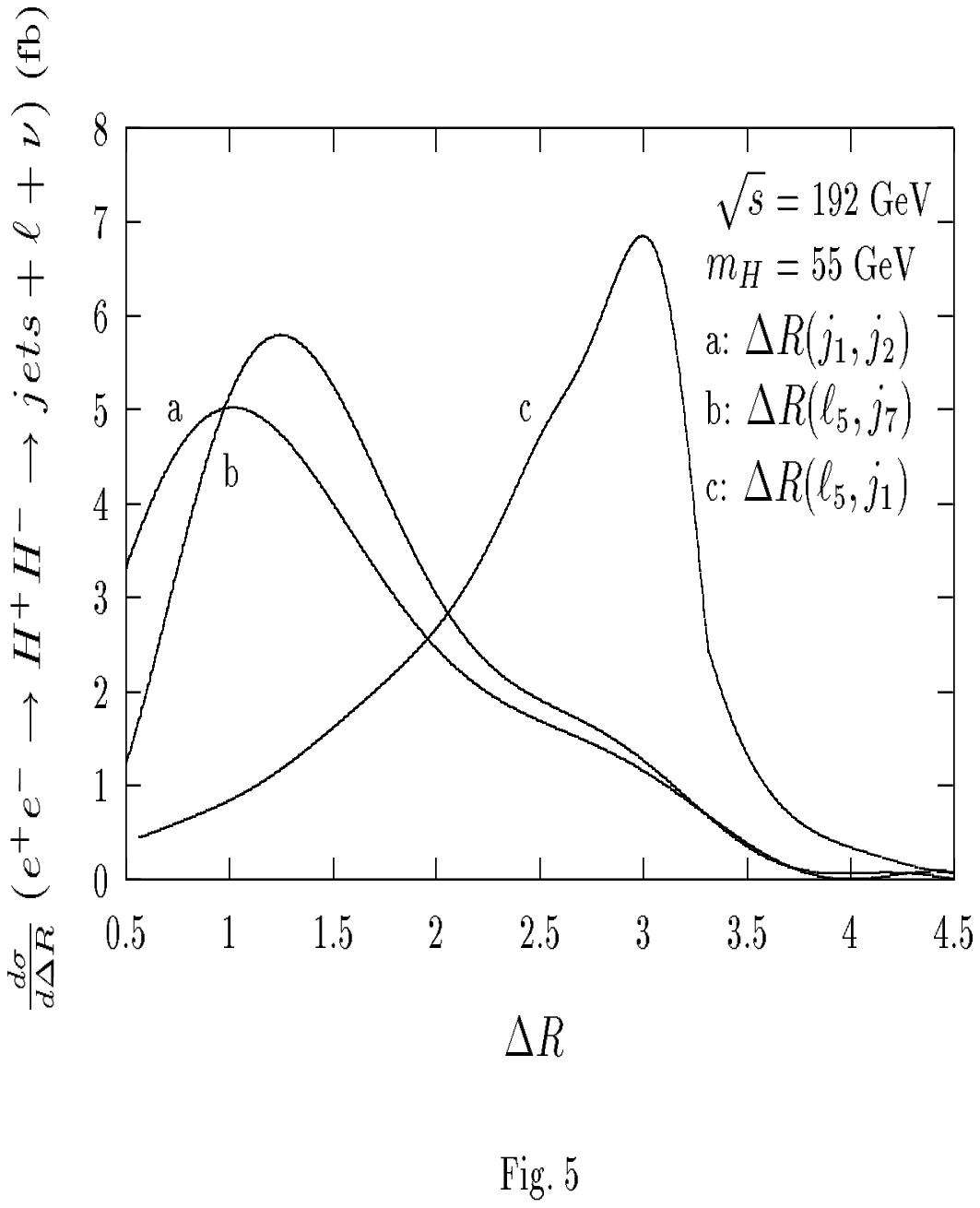}
\end{figure}

\newpage

\begin{figure}[htb]
\epsffile[100 390 120 700]{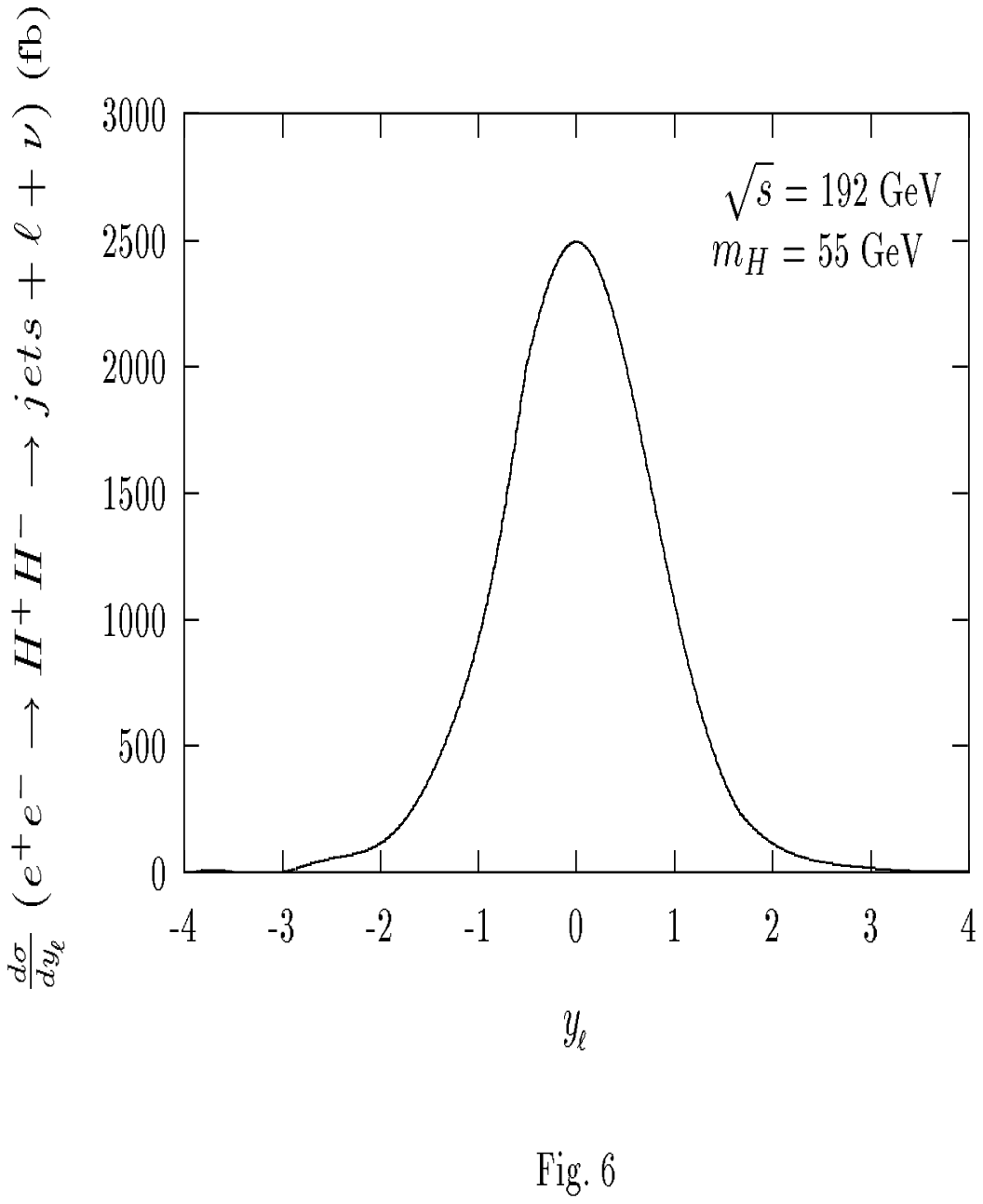}
\end{figure}

\newpage

\begin{figure}[htb]
\epsffile[100 390 120 700]{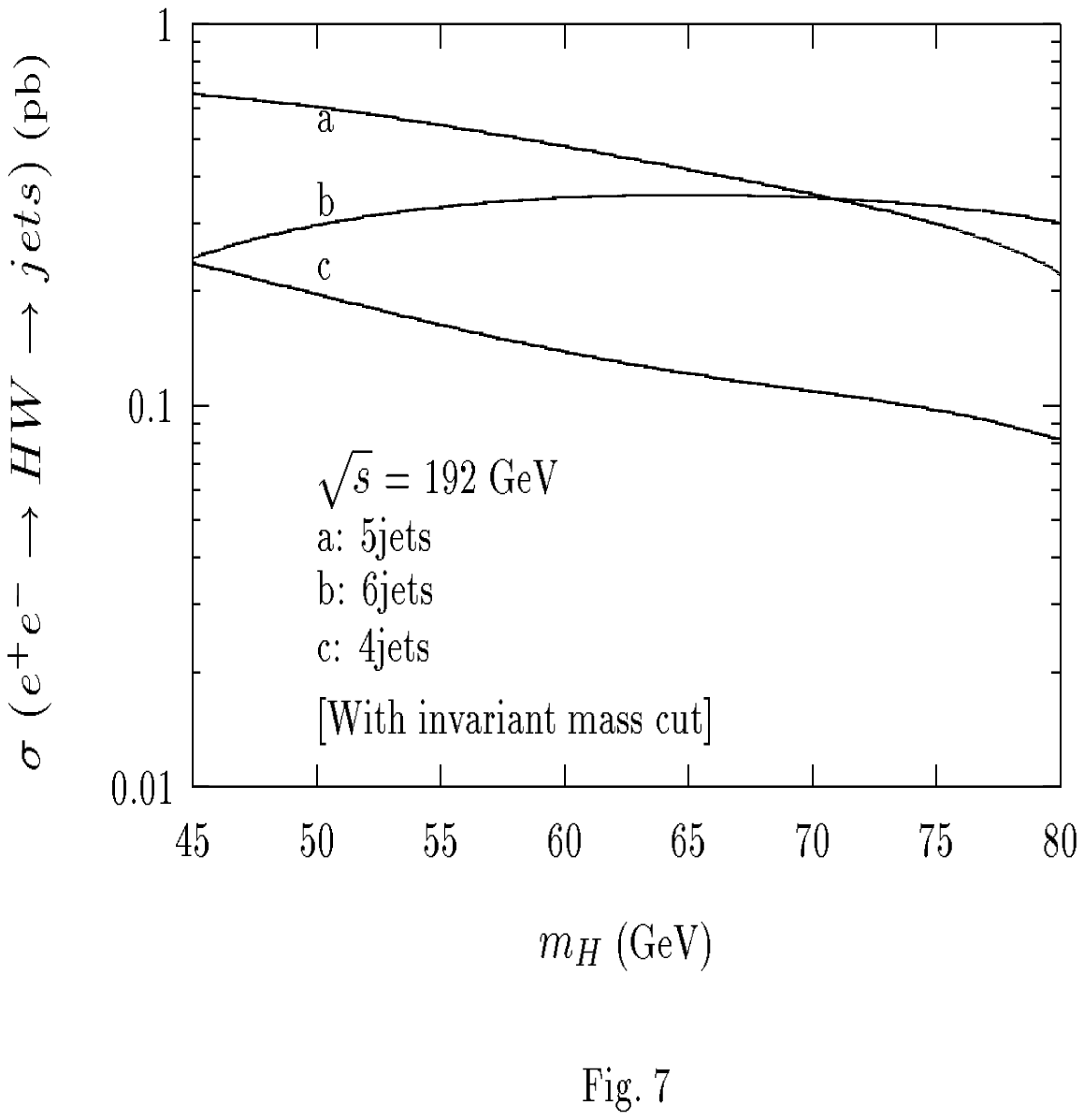}
\end{figure}

\end{document}